# NON-EUCLIDEAN STATISTICS FOR COVARIANCE MATRICES, WITH APPLICATIONS TO DIFFUSION TENSOR IMAGING[1]

By Ian L. Dryden, Alexey Koloydenko and Diwei Zhou

*University of South Carolina, Royal Holloway, University of London and University of Nottingham*

The statistical analysis of covariance matrix data is considered and, in particular, methodology is discussed which takes into account the non-Euclidean nature of the space of positive semi-definite symmetric matrices. The main motivation for the work is the analysis of diffusion tensors in medical image analysis. The primary focus is on estimation of a mean covariance matrix and, in particular, on the use of Procrustes size-and-shape space. Comparisons are made with other estimation techniques, including using the matrix logarithm, matrix square root and Cholesky decomposition. Applications to diffusion tensor imaging are considered and, in particular, a new measure of fractional anisotropy called Procrustes Anisotropy is discussed.

**1. Introduction.** The statistical analysis of covariance matrices occurs in many important applications, for example, in diffusion tensor imaging [Alexander (2005); Schwartzman, Dougherty and Taylor (2008)] or longitudinal data analysis [Daniels and Pourahmadi (2002)]. We wish to consider the situation where the data at hand are sample covariance matrices, and we wish to estimate the population covariance matrix and carry out statistical inference. An example application is in diffusion tensor imaging where a diffusion tensor is a covariance matrix related to the molecular displacement at a particular voxel in the brain, as described in Section 2.

If a sample of covariance matrices is available, we wish to estimate an average covariance matrix, or we may wish to interpolate in space between two or more estimated covariance matrices, or we may wish to carry out tests for equality of mean covariance matrices in different groups.

---

Received May 2008; revised March 2009.
[1]Supported by a Leverhulme Research Fellowship and a Marie Curie Research Training award.
*Key words and phrases.* Anisotropy, Cholesky, geodesic, matrix logarithm, principal components, Procrustes, Riemannian, shape, size, Wishart.

The usual approach to estimating mean covariance matrices in Statistics is to assume a scaled Wishart distribution for the data, and then the maximum likelihood estimator (m.l.e.) of the population covariance matrix is the arithmetic mean of the sample covariance matrices. The estimator can be formulated as a least squares estimator using Euclidean distance. However, since the space of positive semi-definite symmetric matrices is a non-Euclidean space, it is more natural to use alternative distances. In Section 3 we define what is meant by a mean covariance matrix in a non-Euclidean space, using the Fréchet mean. We then review some recently proposed techniques based on matrix logarithms and also consider estimators based on matrix decompositions, such as the Cholesky decomposition and the matrix square root.

In Section 4 we introduce an alternative approach to the statistical analysis of covariance matrices using the Kendall's (1989) size-and-shape space. Distances, minimal geodesics, sample Fréchet means, tangent spaces and practical estimators based on Procrustes analysis are all discussed. We investigate properties of the estimators, including consistency.

In Section 5 we compare the various choices of metrics and their properties. We investigate measures of anisotropy and discuss the deficient rank case in particular. We consider the motivating applications in Section 6 where the analysis of diffusion tensor images and a simulation study are investigated. Finally, we conclude with a brief discussion.

**2. Diffusion tensor imaging.** In medical image analysis a particular type of covariance matrix arises in diffusion weighted imaging called a diffusion tensor. The diffusion tensor is a $3 \times 3$ covariance matrix which is estimated at each voxel in the brain, and is obtained by fitting a physically-motivated model on measurements from the Fourier transform of the molecule displacement density [Basser, Mattiello and Le Bihan (1994); Alexander (2005)].

In the diffusion tensor model the water molecules at a voxel diffuse according to a multivariate normal model centered on the voxel and with covariance matrix $\Sigma$. The displacement of a water molecule $x \in \mathbf{R}^3$ has probability density function

$$f(x) = \frac{1}{(2\pi)^{3/2}|\Sigma|^{1/2}} \exp\left(-\frac{1}{2}x^\mathrm{T}\Sigma^{-1}x\right).$$

The convention is to call $D = \Sigma/2$ the diffusion tensor, which is a symmetric positive semi-definite matrix. The diffusion tensor is estimated at each voxel in the image from the available MR images. The MR scanner has a set of magnetic field gradients applied at directions $g_1, g_2, \ldots, g_m \in RP^2$ with scanner gradient parameter $b$, where $RP^2$ is the real projective space of axial directions (with $g_j \equiv -g_j$, $\|g_j\| = 1$). The data at a voxel consist of signals $(Z_0, Z_1, \ldots, Z_m)$ which are related to the Fourier transform of the displacements in axial direction $g_j \in RP^2, j = 1, \ldots, m$, and the reading $Z_0$ is obtained with no gradient ($b = 0$). The Fourier transform in axial direction $g \in RP^2$ of the multivariate Gaussian displacement density is given by

$$\mathcal{F}(g) = \int \exp(i\sqrt{b}g) f(x)\, dx = \exp(-bg^{\mathrm{T}} D g),$$

and the theoretical model for the signals is

$$Z_j = Z_0 \mathcal{F}(g_j) = Z_0 \exp(-b g_j^{\mathrm{T}} D g_j), \qquad j = 1, \ldots, m.$$

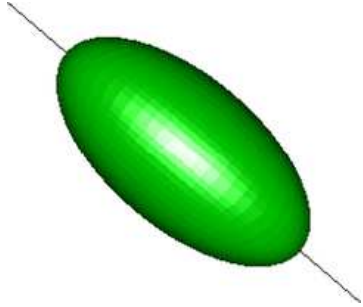

Fig. 1. *Visualization of a diffusion tensor as an ellipsoid. The principal axis is also displayed.*

There are a variety of methods available for estimating $D$ from the data $(Z_0, Z_1, \ldots, Z_m)$ at each voxel [see Alexander (2005)], including least squares regression and Bayesian estimation [e.g., Zhou et al. (2008)]. Noise models include log-Gaussian, Gaussian and, more recently, Rician noise [Wang et al. (2004); Fillard et al. (2007); Basu, Fletcher and Whitaker (2006)]. A common method for visualizing a diffusion tensor is an ellipsoid with principal axes given by the eigenvectors of $D$, and lengths of axes proportional to $\sqrt{\lambda_i}, i = 1, 2, 3$. An example is given in Figure 1.

If a sample of diffusion tensors is available, we may wish to estimate an average diffusion tensor matrix, investigate the structure of variability in diffusion tensors or interpolate at higher spatial resolution between two or more estimated diffusion tensor matrices.

In diffusion tensor imaging a strongly anisotropic diffusion tensor indicates a strong direction of white matter fiber tracts, and plots of measures of anisotropy are very useful to neurologists. A measure that is very commonly used in diffusion tensor imaging is Fractional Anistropy,

$$FA = \left\{ \frac{k}{k-1} \sum_{i=1}^{k} (\lambda_i - \bar{\lambda})^2 \bigg/ \sum_{i=1}^{k} \lambda_i^2 \right\}^{1/2},$$

where $0 \leq FA \leq 1$ and $\lambda_i$ are the eigenvalues of the diffusion tensor matrix. Note that $FA \approx 1$ if $\lambda_1 \gg \lambda_i \approx 0, i > 1$ (very strong principal axis) and $FA = 0$ for isotropy. In diffusion tensor imaging $k = 3$.

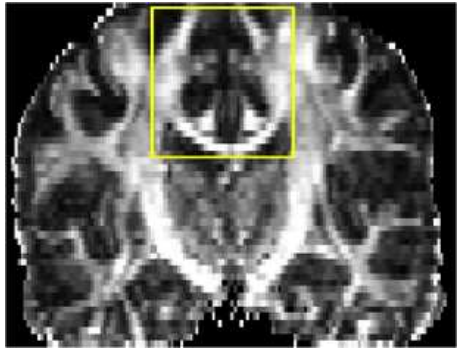

Fig. 2. *An FA map from a slice in a human brain. Lighter values indicate higher FA.*

In Figure 2 we see a plot of FA from an example healthy human brain. We focus on the small inset region in the box, and we would like to interpolate the displayed image to a finer scale. We return to this application in Section 6.3.

## 3. Covariance matrix estimation.

3.1. *Euclidean distance.* Let us consider $n$ sample covariance matrices (symmetric and positive semi-definite $k \times k$ matrices) $S_1, \ldots, S_n$ which are our data (or sufficient statistics). We assume that the $S_i$ are independent and identically distributed (i.i.d.) from a distribution with mean covariance matrix $\Sigma$, although we shall elaborate more later in Section 3.2 about what is meant by a "mean covariance matrix." The main aim is to estimate $\Sigma$. More complicated modeling scenarios are also of interest, but for now we just concentrate on estimating the mean covariance matrix $\Sigma$.

The most common approach is to assume i.i.d. scaled Wishart distributions for $S_i$ with $E(S_i) = \Sigma$, and the m.l.e. for $\Sigma$ is $\hat{\Sigma}_E = \frac{1}{n}\sum_{i=1}^{n} S_i$. This estimator can also be obtained if using a least squares approach by minimizing the sum of square Euclidean distances. The Euclidean distance between two matrices is given by

$$(1) \qquad d_E(S_1, S_2) = \|S_1 - S_2\| = \sqrt{\operatorname{trace}\{(S_1 - S_2)^{\mathrm{T}}(S_1 - S_2)\}},$$

where $\|X\| = \sqrt{\operatorname{trace}(X^{\mathrm{T}} X)}$ is the Euclidean norm (also known as the Frobenius norm). The least squares estimator is given by

$$\hat{\Sigma}_E = \arg\inf_{\Sigma} \sum_{i=1}^{n} \|S_i - \Sigma\|^2.$$

However, the space of positive semi-definite symmetric matrices is a non-Euclidean space and other choices of distance are more natural. One particular drawback with Euclidean distance is when extrapolating beyond the data, nonpositive semi-definite estimates can be obtained. There are other drawbacks when interpolating covariance matrices, as we shall see in our applications in Section 6.

3.2. *The Fréchet mean.* When using a non-Euclidean distance $d(\cdot)$ we must define what is meant by a "mean covariance matrix." Consider a probability distribution for a $k \times k$ covariance matrix $S$ on a Riemannian metric space with density $f(S)$. The Fréchet (1948) mean $\Sigma$ is defined as

$$\Sigma = \arg\inf_{\Sigma} \frac{1}{2} \int d(S, \Sigma)^2 f(S)\, dS,$$

and is also known as the Karcher mean [Karcher (1977)]. The Fréchet mean need not be unique in general, although for many distributions it will be. Provided the distribution is supported only on the geodesic ball of radius $r$, such that the geodesic ball of radius $2r$ is regular [i.e., supremum of sectional curvatures is less than $(\pi/(2r))^2$], then the Fréchet mean $\Sigma$ is unique [Le (1995)]. The support to ensure uniqueness can be very large. For example, for Euclidean spaces (with sectional curvature zero), or for non-Euclidean spaces with negative sectional curvature, the Fréchet mean is always unique.

If we have a sample $S_1, \ldots, S_n$ of i.i.d. observations available, then the sample Fréchet mean is calculated by finding

$$\hat{\Sigma} = \arg\inf_{\Sigma} \sum_{i=1}^{n} d(S_i, \Sigma)^2.$$

Uniqueness of the sample Fréchet mean can also be determined from the result of Le (1995).

3.3. *Non-Euclidean covariance estimators.* A recently derived approach to covariance matrix estimation is to use matrix logarithms. We write the logarithm of a positive definite covariance matrix $S$ as follows. Let $S = U\Lambda U^{\mathrm{T}}$ be the usual spectral decomposition, with $U \in O(k)$ an orthogonal matrix and $\Lambda$ diagonal with strictly positive entries. Let $\log \Lambda$ be a diagonal matrix with logarithm of the diagonal elements of $\Lambda$ on the diagonal. The logarithm of $S$ is given by $\log S = U(\log \Lambda)U^{\mathrm{T}}$ and likewise the exponential of the matrix $S$ is $\exp S = U(\exp \Lambda)U^{\mathrm{T}}$. Arsigny et al. (2007) propose the use of the log-Euclidean distance, where Euclidean distance between the logarithm of covariance matrices is used for statistical analysis, that is,

(2) $$d_L(S_1, S_2) = \|\log(S_1) - \log(S_2)\|.$$

An estimator for the mean population covariance matrix using this approach is given by

$$\hat{\Sigma}_L = \exp\left\{\arg\inf_{\Sigma} \sum_{i=1}^{n} \|\log S_i - \log \Sigma\|^2\right\} = \exp\left\{\frac{1}{n}\sum_{i=1}^{n} \log S_i\right\}.$$

Using this metric avoids extrapolation problems into matrices with negative eigenvalues, but it cannot deal with positive semi-definite matrices of deficient rank.

A further logarithm-based estimator uses a Riemannian metric in the space of square symmetric positive definite matrices

$$(3) \qquad d_R(S_1, S_2) = \|\log(S_1^{-1/2} S_2 S_1^{-1/2})\|.$$

The estimator (sample Fréchet mean) is given by

$$\hat{\Sigma}_R = \arg\inf_{\Sigma} \sum_{i=1}^n \|\log(S_i^{-1/2} \Sigma S_i^{-1/2})\|^2,$$

which has been explored by Pennec, Fillard and Ayache (2006), Moakher (2005), Schwartzman (2006), Lenglet, Rousson and Deriche (2006) and Fletcher and Joshi (2007). The estimate can be obtained using a gradient descent algorithm [e.g., see Pennec (1999); Pennec, Fillard and Ayache (2006)]. Note that this Riemannian metric space has negative sectional curvature and so the population and sample Fréchet means are unique in this case.

Alternatively, one can use a reparameterization of the covariance matrix, such as the Cholesky decomposition [Wang et al. (2004)], where $S_i = L_i L_i^{\mathrm{T}}$ and $L_i = \mathrm{chol}(S_i)$ is lower triangular with positive diagonal entries. The Cholesky distance is given by

$$(4) \qquad d_C(S_1, S_2) = \|\mathrm{chol}(S_1) - \mathrm{chol}(S_2)\|.$$

A least squares estimator can be obtained from

$$\hat{\Sigma}_C = \hat{\Delta}_C \hat{\Delta}_C^{\mathrm{T}}, \qquad \text{where } \hat{\Delta}_C = \arg\inf_{\Delta} \left\{ \frac{1}{n} \sum_{i=1}^n \|L_i - \Delta\|^2 \right\} = \frac{1}{n} \sum_{i=1}^n L_i.$$

An equivalent model-based approach would use an independent Gaussian perturbation model for the lower triangular part of $L_i$, with mean given by the lower triangular part of $\Delta_C$, and so $\hat{\Delta}_C$ is the m.l.e. of $\Delta_C$ under this model. Hence, in this approach the averaging is carried out on a square root type-scale, which would indeed be the case for $k = 1$ dimensional case where the estimate of variance would be the square of the mean of the sample standard deviations.

An alternative decomposition is the matrix square root where $S^{1/2} = U\Lambda^{1/2} U^{\mathrm{T}}$, which has not been used in this context before as far as we are aware. The distance is given by

$$(5) \qquad d_H(S_1, S_2) = \|S_1^{1/2} - S_2^{1/2}\|.$$

A least squares estimator can be obtained from

$$\hat{\Sigma}_H = \hat{\Delta}_H \hat{\Delta}_H^{\mathrm{T}}, \qquad \text{where } \hat{\Delta}_H = \arg\inf_{\Delta} \left\{ \sum_{i=1}^n \|S_i^{1/2} - \Delta\|^2 \right\} = \frac{1}{n} \sum_{i=1}^n S_i^{1/2}.$$

However, because $L_i R R^{\mathrm{T}} L_i^{\mathrm{T}} = L_i L_i^{\mathrm{T}}$ for $R \in O(k)$, another new alternative is to relax the lower triangular or square root parameterizations and match the initial decompositions closer in terms of Euclidean distance by optimizing over rotations and reflections. This idea provides the rationale for the main approaches in this paper.

## 4. Procrustes size-and-shape analysis.

4.1. *Non-Euclidean size-and-shape metric.* The non-Euclidean size-and-shape metric between two $k \times k$ covariance matrices $S_1$ and $S_2$ is defined as

(6) $$d_S(S_1, S_2) = \inf_{R \in O(k)} \|L_1 - L_2 R\|,$$

where $L_i$ is a decomposition of $S_i$ such that $S_i = L_i L_i^{\mathrm{T}}, i = 1, 2$. For example, we could have the Cholesky decomposition $L_i = \mathrm{chol}(S_i), i = 1, 2$, which is lower triangular with positive diagonal elements, or we could consider the matrix square root $L = S^{1/2} = U\Lambda^{1/2}U^{\mathrm{T}}$, where $S = U\Lambda U^{\mathrm{T}}$ is the spectral decomposition. Note that $S_1 = (L_1 R)(L_1 R)^{\mathrm{T}}$ for any $R \in O(k)$, and so the distance involves matching $L_1$ optimally, in a least-squares sense, to $L_2$ by rotation and reflection. Since $S = LL^{\mathrm{T}}$, then the decomposition is represented by an equivalence class $\{LR : R \in O(k)\}$. For practical computation we often need to choose a representative from this class, called an icon, and in our computations we shall choose the Cholesky decomposition.

The Procrustes solution for matching $L_2$ to $L_1$ is

(7) $$\begin{aligned}\hat{R} &= \arg\inf_{R \in O(k)} \|L_1 - L_2 R\| \\ &= UW^{\mathrm{T}}, \qquad \text{where } L_1^{\mathrm{T}} L_2 = W\Lambda U^{\mathrm{T}}, U, W \in O(k),\end{aligned}$$

and $\Lambda$ is a diagonal matrix of positive singular values [e.g., see Mardia, Kent and Bibby (1979), page 416].

This metric has been used previously in the analysis of point set configurations where invariance under translation, rotation and reflection is required. Size-and-shape spaces were introduced by Le (1988) and Kendall (1989) as part of the pioneering work on the shape analysis of landmark data [cf. Kendall (1984)]. The detailed geometry of these spaces is given by Kendall et al. [(1999), pages 254–264], and, in particular, the size-and-shape space is a cone with a warped-product metric and has positive sectional curvature.

Equation (6) is a Riemannian metric in the reflection size-and-shape space of $(k+1)$-points in $k$ dimensions [Dryden and Mardia (1998), Chapter 8]. In particular, $d_S(\cdot)$ is the reflection size-and-shape distance between the $(k+1) \times k$ configurations $H^T L_1$ and $H^T L_2$, where $H$ is the $k \times (k+1)$ Helmert sub-matrix [Dryden and Mardia (1998), page 34] which has $j$th row given by

$$(\underbrace{h_j, \ldots, h_j}_{j \text{ times}}, -jh_j, \underbrace{0, \ldots, 0}_{k-j \text{ times}}), \qquad h_j = -\{j(j+1)\}^{-1/2},$$

for $j = 1, \ldots, k$.

Hence, the statistical analysis of covariance matrices can be considered equivalent to the dual problem of analyzing reflection size-and-shapes.

4.2. *Minimal geodesic and tangent space.* Let us consider the minimal geodesic path through the reflection size-and-shapes of $L_1$ and $L_2$ in the reflection size-and-shape space, where $L_i L_i^T = S_i$, $i = 1, 2$. Following an argument similar to that for the minimal geodesics in shape spaces [Kendall et al. (1999)], this minimal geodesic can be isometrically expressed as $L_1 + tT$, where $T$ are the horizontal tangent co-ordinates of $L_2$ with pole $L_1$. Kendall et al. [(1999), Section 11.2] discuss size-and-shape spaces without reflection invariance, however, the results with reflection invariance are similar, as reflection does not change the local geometry.

The horizontal tangent coordinates satisfy $L_1 T^T = T L_1^T$ [Kendall et al. (1999), page 258]. Explicitly, the horizontal tangent coordinates are given by

$$T = L_2 \hat{R} - L_1, \qquad \hat{R} = \inf_{R \in O(k)} \|L_1 - L_2 R\|,$$

where $\hat{R}$ is the Procrustes match of $L_2$ onto $L_1$ given in (7). So, the geodesic path starting at $L_1$ and ending at $L_2$ is given by

$$w_1 L_1 + w_2 L_2 \hat{R},$$

where $w_1 + w_2 = 1, w_i \geq 0, i = 1, 2$, and $\hat{R}$ is given in (7). Minimal geodesics are useful in applications for interpolating between two covariance matrices, in regression modeling of a series of covariance matrices, and for extrapolation and prediction.

Tangent spaces are very useful in practical applications, where one uses Euclidean distances in the tangent space as approximations to the non-Euclidean metrics in the size-and-shape space itself. Such constructions are useful for approximate multivariate normal based inference, dimension reduction using principal components analysis and large sample asymptotic distributions.

4.3. *Procrustes mean covariance matrix.* Let $S_1, \ldots, S_n$ be a sample of $n$ positive semi-definite covariance matrices each of size $k \times k$ from a distribution with density $f(S)$, and we work with the Procrustes metric (6) in order to estimate the Fréchet mean covariance matrix $\Sigma$. We assume that $f(S)$ leads to a unique Fréchet mean (see Section 3.2).

The sample Fréchet mean is calculated by finding

$$\hat{\Sigma}_S = \arg\inf_{\Sigma} \sum_{i=1}^{n} d_S(S_i, \Sigma)^2.$$

In the dual size-and-shape formulation we can write this as

(8) $\quad \hat{\Sigma}_S = \hat{\Delta}_S \hat{\Delta}_S^{\mathrm{T}}, \qquad \text{where } \hat{\Delta}_S = \arg\inf_{\Delta} \sum_{i=1}^{n} \inf_{R_i \in O(k)} \|H^{\mathrm{T}} L_i R_i - H^{\mathrm{T}} \Delta\|^2.$

The solution can be found using the Generalized Procrustes Algorithm [Gower (1975); Dryden and Mardia (1998), page 90], which is available in the shapes library (written by the first author of this paper) in R [R Development Core Team (2007)]. Note that if the data lie within a geodesic ball of radius $r$ such that the geodesic ball of radius $2r$ is regular [Le (1995); Kendall (1990)], then the algorithm finds the global unique minimum solution to (8). This condition can be checked for any dataset and, in practice, the algorithm works very well indeed.

4.4. *Tangent space inference.* If the variability in the data is not too large, then we can project the data into the tangent space and carry out the usual Euclidean based inference in that space.

Consider a sample $S_1, \ldots, S_n$ of covariance matrices with sample Fréchet mean $\hat{\Sigma}_S$ and tangent space coordinates with pole $\hat{\Sigma}_S = \hat{\Delta}_S \hat{\Delta}_S^{\mathrm{T}}$ given by

$$V_i = \hat{\Delta}_S - L_i \hat{R}_i,$$

where $\hat{R}_i$ is the Procrustes rotation for matching $L_i$ to $\hat{\Delta}_S$, $i = 1, \ldots, n$, and $S_i = L_i L_i^{\mathrm{T}}, i = 1, 2$.

Frequently one wishes to reduce the dimension of the problem, for example, using principal components analysis. Let

$$S_v = \frac{1}{n} \sum_{i=1}^{n} \mathrm{vec}(V_i) \mathrm{vec}(V_i)^{\mathrm{T}},$$

where vec is the vectorize operation. The principal component (PC) loadings are given by $\hat{\gamma}_j, j = 1, \ldots, p$, the eigenvectors of $S_v$ corresponding to the eigenvalues $\hat{\lambda}_1 \geq \hat{\lambda}_2 \geq \cdots \geq \hat{\lambda}_p > 0$, where $p$ is the number of nonzero eigenvalues. The PC score for the $i$th individual on PC $j$ is given by

$$s_{ij} = \hat{\gamma}_j^{\mathrm{T}} \mathrm{vec}(V_i), \qquad i = 1, \ldots, n, j = 1, \ldots, p.$$

In general, $p = \min(n - 1, k(k + 1)/2)$. The effect of the $j$th PC can be examined by evaluating

$$\Sigma(c) = (\hat{\Delta}_S + c\operatorname{vec}_k^{-1}(\hat{\lambda}_j^{1/2}\hat{\gamma}_j))(\hat{\Delta}_S + c\operatorname{vec}_k^{-1}(\hat{\lambda}_j^{1/2}\hat{\gamma}_j))^{\mathrm{T}}$$

for various $c$ [often in the range $c \in (-3, 3)$, for example], where $\operatorname{vec}_k^{-1}(\operatorname{vec}(V)) = V$ for a $k \times k$ matrix $V$.

Tangent space inference can proceed on the first $p$ PC scores, or possibly in lower dimensions if desired. For example, Hotelling's $T^2$ test can be carried out to examine group differences, or regression models could be developed for investigating the PC scores as responses versus various covariates. We shall consider principal components analysis of covariance matrices in an application in Section 6.2.

4.5. *Consistency.* Le (1995, 2001) and Bhattacharya and Patrangenaru (2003, 2005) provide consistency results for Riemannian manifolds, which can be applied directly to our situation. Consider a distribution $F$ on the space of covariance matrices which has size-and-shape Fréchet mean $\Sigma_S$. Let $S_1, \ldots, S_n$ be i.i.d. from $F$, such that they lie within a geodesic ball $B_r$ such that $B_{2r}$ is regular. Then

$$\hat{\Sigma}_S \xrightarrow{P} \Sigma_S, \qquad \text{as } n \to \infty,$$

where $\Sigma_S$ is unique. In addition, we can derive a central limit theorem result as in Bhattacharya and Patrangenaru (2005), where the tangent coordinates have an approximate multivariate normal distribution for large $n$. Hence, confidence regions based on the bootstrap can be obtained, as in Amaral, Dryden and Wood (2007) and Bhattacharya and Patrangenaru (2003, 2005).

4.6. *Scale invariance.* In some applications it may be of interest to consider invariance over isotropic scaling of the covariance matrix. In this case we could consider the representation of the covariance matrix using Kendall's reflection shape space, with the shape metric given by the full Procrustes distance

$$(9) \qquad d_F(S_1, S_2) = \inf_{R \in O(k), \beta > 0} \left\| \frac{L_1}{\|L_1\|} - \beta L_2 R \right\|,$$

where $S_i = L_i L_i^{\mathrm{T}}, i = 1, 2$, and $\beta > 0$ is a scale parameter. Another choice of the estimated covariance matrix from a sample $S_1, \ldots, S_n$, which is scale invariant and based on the full Procrustes mean shape (extrinsic mean), is

$$\hat{\Sigma}_F = \hat{\Delta}_F \hat{\Delta}_F^{\mathrm{T}}, \qquad \text{where } \hat{\Delta}_F = \arg\inf_{\Delta} \sum_{i=1}^{n} \Big\{ \inf_{R_i \in O(k),\ \beta_i > 0} \|\beta_i L_i R_i - \Delta\|^2 \Big\},$$

TABLE 1
*Notation and definitions of the distances and estimators*

| Name | Notation | Form | Estimator | Equation |
|---|---|---|---|---|
| Euclidean | $d_E(S_1, S_2)$ | $\|S_1 - S_2\|$ | $\hat{\Sigma}_E$ | (1) |
| Log-Euclidean | $d_L(S_1, S_2)$ | $\|\log(S_1) - \log(S_2)\|$ | $\hat{\Sigma}_L$ | (2) |
| Riemannian | $d_R(S_1, S_2)$ | $\|\log(S_1^{-1/2} S_2 S_1^{-1/2})\|$ | $\hat{\Sigma}_R$ | (3) |
| Cholesky | $d_C(S_1, S_2)$ | $\|\text{chol}(S_1) - \text{chol}(S_2)\|$ | $\hat{\Sigma}_C$ | (4) |
| Root Euclidean | $d_H(S_1, S_2)$ | $\|S_1^{1/2} - S_2^{1/2}\|$ | $\hat{\Sigma}_H$ | (5) |
| Procrustes size-and-shape | $d_S(S_1, S_2)$ | $\inf_{R \in O(k)} \|L_1 - L_2 R\|$ | $\hat{\Sigma}_S$ | (6) |
| Full Procrustes shape | $d_F(S_1, S_2)$ | $\inf_{R \in O(k), \beta > 0} \|\frac{L_1}{\|L_1\|} - \beta L_2 R\|$ | $\hat{\Sigma}_F$ | (9) |
| Power Euclidean | $d_A(S_1, S_2)$ | $\frac{1}{\alpha}\|S_1^\alpha - S_2^\alpha\|$ | $\hat{\Sigma}_A$ | (10) |

and $S_i = L_i L_i^T, i = 1, \ldots, n$, and $\beta_i > 0$ are scale parameters. The solution can again be found from the Generalized Procrustes Algorithm using the shapes library in R. Tangent space inference can then proceed in an analogous manner to that of Section 4.4.

## 5. Comparison of approaches.

5.1. *Choice of metrics.* In applications there are several choices of distances between covariance matrices that one could consider. For completeness we list the metrics and the estimators considered in this paper in Table 1, and we discuss briefly some of their properties.

Estimators $\hat{\Sigma}_E, \hat{\Sigma}_C, \hat{\Sigma}_H, \hat{\Sigma}_L, \hat{\Sigma}_A$ are straightforward to compute using arithmetic averages. The Procrustes based estimators $\hat{\Sigma}_S, \hat{\Sigma}_F$ involve the use of the Generalized Procrustes Algorithm, which works very well in practice. The Riemannian metric estimator $\hat{\Sigma}_R$ uses a gradient descent algorithm which is guaranteed to converge [see Pennec (1999); Pennec, Fillard and Ayache (2006)].

All these distances except $d_C$ are invariant under simultaneous rotation and reflection of $S_1$ and $S_2$, that is, the distances are unchanged by replacing both $S_i$ by $VS_i V^T, V \in O(k), i = 1, 2$. Metrics $d_L(\cdot), d_R(\cdot), d_F(\cdot)$ are invariant under simultaneous scaling of $S_i, i = 1, 2$, that is, replacing both $S_i$ by $\beta S_i$, $\beta > 0$. Metric $d_R(\cdot)$ is also affine invariant, that is, the distances are unchanged by replacing both $S_i$ by $AS_i A^T, i = 1, 2$, where $A$ is a general $k \times k$ full rank matrix. Metrics $d_L(\cdot), d_R(\cdot)$ have the property that

$$d(A, I_k) = d(A^{-1}, I_k),$$

where $I_k$ is the $k \times k$ identity matrix.

Metrics $d_L(\cdot), d_R(\cdot), d_F(\cdot)$ are not valid for comparing rank deficient covariance matrices. Finally, there are problems with extrapolation with metric $d_E(\cdot)$: extrapolate too far and the matrices are no longer positive semi-definite.

5.2. *Anisotropy.* In some applications a measure of anisotropy of the covariance matrix may be required, and in Section 2 we described the commonly used FA measure. An alternative is to use the full Procrustes shape distance to isotropy and we have

$$PA = \sqrt{\frac{k}{k-1}} d_F(I_k, S) = \sqrt{\frac{k}{k-1}} \inf_{R \in O(k), \beta \in \mathbf{R}^+} \left\| \frac{I_k}{\sqrt{k}} - \beta \operatorname{chol}(S) R \right\|,$$

$$= \left\{ \frac{k}{k-1} \sum_{i=1}^{k} (\sqrt{\lambda_i} - \overline{\sqrt{\lambda}})^2 \Big/ \sum_{i=1}^{k} \lambda_i \right\}^{1/2},$$

where $\overline{\sqrt{\lambda}} = \frac{1}{k} \sum \sqrt{\lambda_i}$. Note that the maximal value of $d_F$ distance from isotropy to the rank 1 covariance matrix is $\sqrt{(k-1)/k}$, which follows from Le (1992). We include the scale factor when defining the Procrustes Anisotropy (PA), and so $0 \le PA \le 1$, with $PA = 0$ indicating isotropy, and $PA \approx 1$ indicating a very strong principal axis.

A final measure based on metrics $d_L$ or $d_R$ is the geodesic anisotropy

$$GA = \left\{ \sum_{i=1}^{k} (\log \lambda_i - \overline{\log \lambda})^2 \right\}^{1/2},$$

where $0 \le GA < \infty$ [Arsigny et al. (2007); Fillard et al. (2007); Fletcher and Joshi (2007)], which has been used in diffusion tensor analysis in medical imaging with $k = 3$.

5.3. *Deficient rank case.* In some applications covariance matrices are close to being deficient in rank. For example, when *FA* or *PA* are equal to 1, then the covariance matrix is of rank 1. The Procrustes metrics can easily deal with deficient rank matrices, which is a strong advantage of the approach. Indeed, Kendall's (1984, 1989) original motivation for developing his theory of shape was to investigate rank 1 configurations in the context of detecting "flat" (collinear) triangles in archeology.

The use of $\hat{\Sigma}_L$ and $\hat{\Sigma}_R$ has strong connections with the use of Bookstein's (1986) hyperbolic shape space and Le and Small's (1999) simplex shape space, and such spaces cannot deal with deficient rank configurations.

The use of the Cholesky decomposition has strong connections with Bookstein coordinates and Goodall–Mardia coordinates in shape analysis, where one registers configurations on a common baseline [Bookstein (1986); Goodall and Mardia (1992)]. For small variability the baseline registration method and Procrustes superimposition techniques are similar, and there is an approximate linear relationship between the two [Kent (1994)]. In shape analysis edge superimposition techniques can be very unreliable if the baseline is very small in length, which would correspond to very small variability in particular diagonal elements of the covariance matrix in the current context. Cholesky methods would be unreliable in such cases. Also, Bookstein coordinates induce correlations in the shape variables and, hence, estimation of covariance structure is biased [Kent (1994)]. Hence, in general, Procrustes techniques are preferred over edge superimposition techniques in shape analysis. Hence, this would mean in the current context that the Procrustes approaches of this paper should be preferred to inference using the Cholesky decomposition.

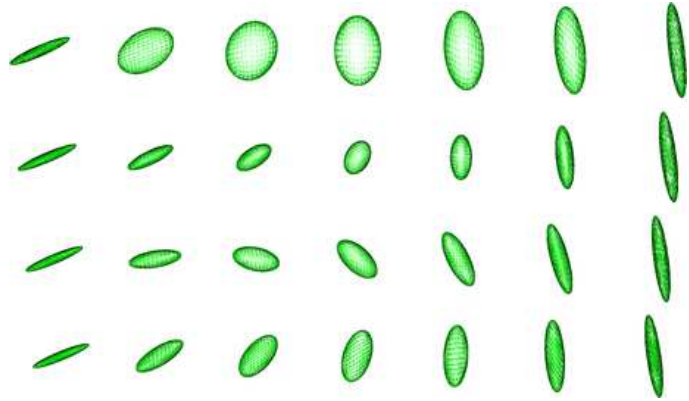

FIG. 3. *Four different geodesic paths between the two tensors. The geodesic paths are obtained using $d_E(\cdot)$ (1st row), $d_L(\cdot)$ (2nd row), $d_C(\cdot)$ (3rd row) and $d_S(\cdot)$ (4th row).*

## 6. Applications.

6.1. *Interpolation of covariance matrices.* Frequently in diffusion tensor imaging one wishes to carry out interpolation between tensors. When the tensors are quite different, interpolation using different metrics can lead to very different results. For example, consider Figure 3, where four different geodesic paths are plotted between two tensors. Arsigny et al. (2007) note that the Euclidean metric is prone to swelling, which is seen in this example. Also, the log-Euclidean metric gives strong weight to small volumes. In this example the Cholesky and Procrustes size-and-shape paths look rather different, due to the extra rotation in the Procrustes method. From a variety of examples it does seem clear that the Euclidean metric is very problematic, especially due to the swelling of the volume. In general, the log-Euclidean and Procrustes size-and-shape methods seem preferable.

In some applications, for example, fiber tracking, we may need to interpolate between several covariance matrices on a grid, in which case we can use weighted Fréchet means

$$\hat{\Sigma} = \arg\inf_{\Sigma} \sum_{i=1}^{n} w_i d(S_i, \Sigma)^2, \qquad \sum_{i=1}^{n} w_i = 1,$$

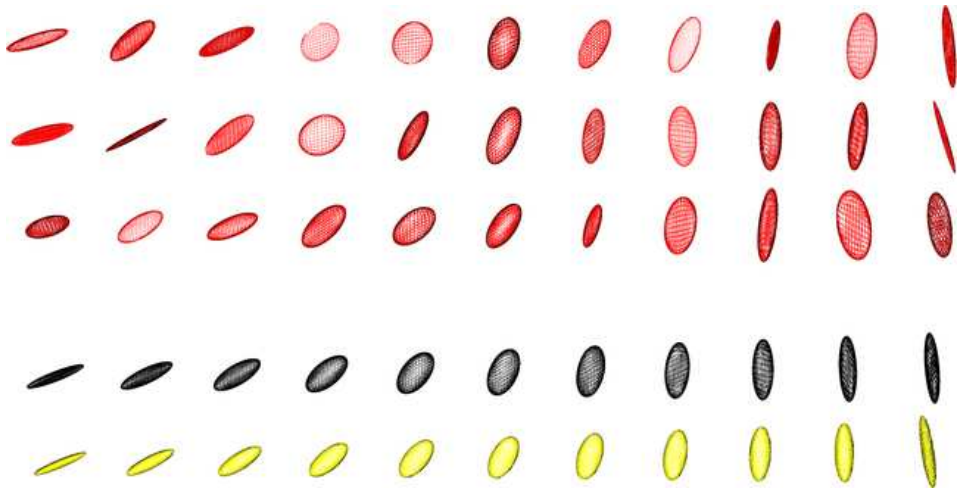

Fig. 4. *Demonstration of PCA for covariance matrices. The true geodesic path is given in the penultimate row (black). We then add noise in the three initial rows (red). Then we estimate the mean and find the first principal component (yellow), displayed in the bottom row.*

where the weights $w_i$ are proportional to a function of the distance (e.g., inverse distance or Kriging based weights).

6.2. *Principal components analysis of diffusion tensors.* We consider now an example estimating the principal geodesics of the covariance matrices $S_1, \ldots, S_n$ using the Procrustes size-and-shape metric. The data are displayed in Figure 4 and here $k = 3$. We consider a true geodesic path (black) and evaluate 11 equally spaced covariance matrices along this path. We then add noise for three separate realizations of noisy paths (in red). The noise is independent and identically distributed Gaussian and is added in the dual space of the tangent coordinates. First, the overall Procrustes size-and-shape mean $\hat{\Sigma}_S$ is computed based on all the data ($n = 33$), and then the Procrustes size-and-shape tangent space co-ordinates are obtained. The first principal component loadings are computed and projected back to give an estimated minimal geodesic in the covariance matrix space. We plot this path in yellow by displaying 11 covariance matrices along the path. As we would expect, the first principal component path bears a strong similarity to the true geodesic path. The percentages of variability explained by the first three PCs are as follows: PC1 (72.0%), PC2 (8.8%), PC3 (6.5%).

The data can also be seen in the dual Procrustes space of 4 points in $k = 3$ dimensions in Figure 5. We also see the data after applying the Procrustes fitting, we show the effects of the first three principal components, and also the plot of the first three PC scores.

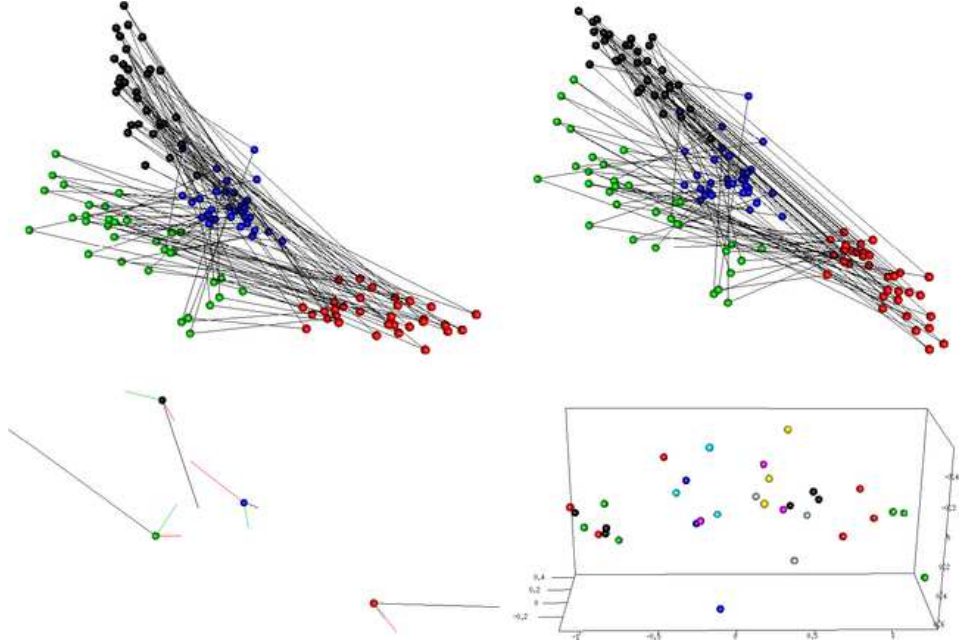

FIG. 5. *(top left) The noisy configurations in the dual space of $k+1=4$ points in $k=3$ dimensions. For each configuration point 1 is colored black, point 2 is red, point 3 is green and point 4 is blue, and the points in a configuration are joined by lines. (top right) The Procrustes registered data, after removing translation, rotation and reflection. (bottom left) The Procrustes mean size-and-shape, with vectors drawn along the directions of the first three PCs (PC1—black, PC2—red, PC3—green). (bottom right) The first three PC scores. The points are colored by the position along the true geodesic from left to right (black, red, green, blue, cyan, purple, yellow, grey, black, red, green).*

6.3. *Interpolation.* We consider the interpolation of part of the brain image in Figure 2. In Figure 6(a) we see the original FA image, and in Figure 6(b) and (c) we see interpolated images using size-and-shape distance. The interpolation is carried out at two equally spaced points between voxels, and Figure 6(b) shows the FA image from the interpolation and Figure 6(c) shows the PA image. In the bottom right plot of Figure 6 we highlight the selected regions in the box. It is clear that the interpolated images are smoother, and it is clear from the anisotropy maps of the interpolated data that the cingulum (cg) is distinct from the corpus callosum (cc).

6.4. *Anisotropy.* As a final application we consider some diffusion tensors obtained from diffusion weighted images in the brain. In Figure 7 we see a coronal slice from the brain with the $3 \times 3$ tensors displayed. This image is a coronal view of the brain, and the corpus callosum and cingulum can be seen. The diagonal tract on the lower left is the anterior limb of the internal capsule and on the lower right we see the superior fronto-occipital fasciculus.

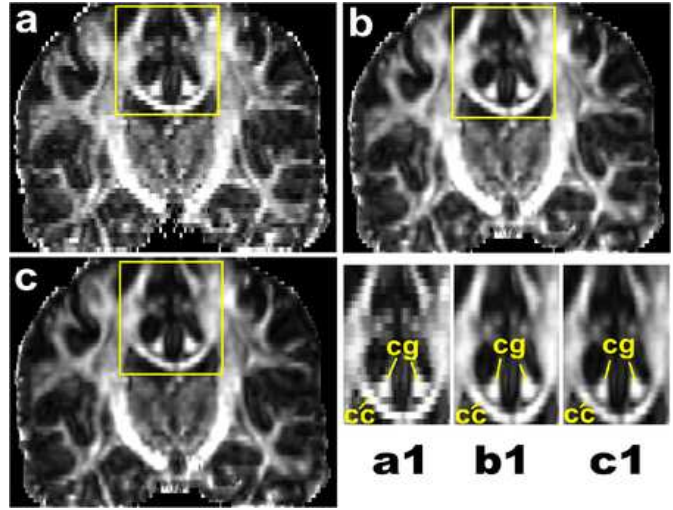

Fig. 6. *FA maps from the original* (a) *and interpolated* (b) *data. In* (c) *the PA map is displayed, and in* (a1), (b1), (c1) *we see the zoomed in regions marked in* (a), (b), (c) *respectively.*

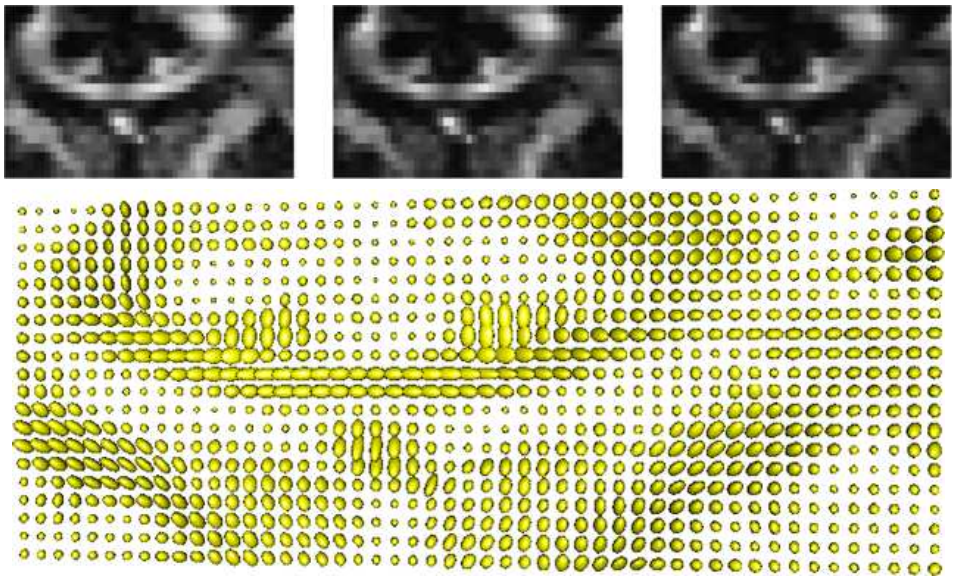

Fig. 7. *In the upper plots we see the anisotropy measures (left) FA, (middle) PA, (right) GA. In the lower plot we see the diffusion tensors, which have been scaled to have volume proportional to $\sqrt{FA}$.*

At first sight all three measures appear broadly similar. However, the PA image offers more contrast than the FA image in the highly anisotropic region—the corpus callosum. Also, the GA image has rather fewer brighter areas than PA or FA. Due to the improved contrast, we believe PA is slightly preferable in this example.

6.5. *Simulation study.* Finally, we consider a simulation study to compare the different estimators. We consider the problem of estimating a population covariance matrix $\Omega$ from a random sample of $k \times k$ covariance matrices $S_1, \ldots, S_n$.

We consider a random sample generated as follows. Let $\Delta = \text{chol}(\Omega)$ and $X_i$ be a random matrix with i.i.d. entries with $E[(X_i)_{jl}] = 0$, $\text{var}((X_i)_{jl}) = \sigma^2$, $i = 1, \ldots, n; j = 1, \ldots, k; l = 1, \ldots, k$. We take

$$S_i = (\Delta + X_i)(\Delta + X_i)^{\text{T}}, \qquad i = 1, \ldots, n.$$

We shall consider four error models:

I. Gaussian square root: $(X_i)_{jl}$ are i.i.d. $N(0, \sigma^2)$ for $j = 1, \ldots, k; l = 1, \ldots, k$.
II. Gaussian Cholesky: $(X_i)_{jl}$ are i.i.d. $N(0, \sigma^2)$ for $j \leq k$ and zero otherwise.
III. Log-Gaussian: i.i.d. Gaussian errors $N(0, \sigma^2)$ are added to the matrix logarithm of $\Delta$ to give $Y$, and then the matrix exponential of $YY^{\text{T}}$ is taken.
IV. Student's $t$ with 3 degrees of freedom: $(X_i)_{jl}$ are i.i.d. $(\sigma/\sqrt{3})t_3$ for $j = 1, \ldots, k; l = 1, \ldots, k$.

We consider the performance in a simulation study, with 1000 Monte Carlo simulations. The results are presented in Tables 2 and 3 for two choices of population covariance matrix. We took $k = 3$ and $n = 10, 30$. In order to investigate the efficiency of the estimators, we use three measures: estimated mean square error between the estimate and the matrix $\Omega$ with metrics $d_E(\cdot), d_S(\cdot)$ and the estimated risk from using Stein loss [James and Stein (1961)] which is given by

$$L(S_1, S_2) = \text{trace}(S_1 S_2^{-1}) - \log \det(S_1 S_2^{-1}) - k,$$

where $\det(\cdot)$ is the determinant. Clearly the efficiency of the methods depends strongly on the $\Omega$ and the error distribution.

Consider the first case where the mean has $\lambda_1 = 1, \lambda_2 = 0.3, \lambda_3 = 0.1$ in Table 2. We discuss model I first where the errors are Gaussian on the matrix square root scale. The efficiency is fairly similar for each estimator for $n = 10$, with $\hat{\Sigma}_H$ performing the best. For $n = 30$ either $\hat{\Sigma}_H$ or $\hat{\Sigma}_S$ are better, with $\hat{\Sigma}_E$ performing least well. For model II with Gaussian errors added in the Cholesky decomposition we see that $\hat{\Sigma}_C$ is the best, although the other estimators are quite similar, with the exception of $\hat{\Sigma}_E$ which is worse. For model III with Gaussian errors on the matrix logarithm scale all estimators are quite similar, as the variability is rather small. The estimate $\hat{\Sigma}_R$ is slightly better here than the others. For model IV with Student's $t_3$ errors we see that $\hat{\Sigma}_H$ and $\hat{\Sigma}_S$ are slightly better on the whole, although $\hat{\Sigma}_E$ is again the worst performer.

TABLE 2
*Measures of efficiency, with $k = 3$ and $\sigma = 0.1$. RMSE is the root mean square error using either the Euclidean norm or the Procrustes size-and-shape norm, and "Stein" refers to the risk using the Stein loss function. The smallest value in each row is highlighted in bold. The mean has parameters $\lambda_1 = 1, \lambda_2 = 0.3, \lambda_3 = 0.1$. The error distributions for Models I–IV are Gaussian (square root), Gaussian (Cholesky), log-Gaussian and Student's $t_3$, respectively*

|  |  | $\hat{\Sigma}_E$ | $\hat{\Sigma}_C$ | $\hat{\Sigma}_S$ | $\hat{\Sigma}_H$ | $\hat{\Sigma}_L$ | $\hat{\Sigma}_R$ | $\hat{\Sigma}_F$ |
|---|---|---|---|---|---|---|---|---|
| I |  |  |  |  |  |  |  |  |
| $n = 10$ | RMSE($d_E$) | 0.1136 | 0.1057 | 0.104 | **0.1025** | 0.104 | 0.1176 | 0.1058 |
|  | RMSE($d_S$) | 0.0911 | 0.082 | 0.0802 | **0.0794** | 0.0851 | 0.0892 | 0.0813 |
|  | Stein | 0.0869 | 0.0639 | 0.0615 | **0.0604** | 0.0793 | 0.0728 | 0.0626 |
| $n = 30$ | RMSE($d_E$) | 0.0788 | 0.0669 | 0.0626 | **0.0611** | 0.0642 | 0.0882 | 0.0652 |
|  | RMSE($d_S$) | 0.0691 | 0.0516 | **0.0475** | 0.0477 | 0.0525 | 0.0607 | 0.049 |
|  | Stein | 0.058 | 0.0242 | **0.0207** | 0.0223 | 0.0295 | 0.0265 | 0.0216 |
| II |  |  |  |  |  |  |  |  |
| $n = 10$ | RMSE($d_E$) | 0.0973 | **0.0889** | 0.0911 | 0.0906 | 0.093 | 0.1014 | 0.0923 |
|  | RMSE($d_S$) | 0.0797 | **0.0695** | 0.0714 | 0.0713 | 0.0752 | 0.0785 | 0.0721 |
|  | Stein | 0.07 | **0.0468** | 0.0499 | 0.0502 | 0.0573 | 0.0554 | 0.0506 |
| $n = 30$ | RMSE($d_E$) | 0.0641 | **0.0513** | 0.0535 | 0.0533 | 0.058 | 0.0732 | 0.0551 |
|  | RMSE($d_S$) | 0.0585 | **0.0399** | 0.0422 | 0.0432 | 0.0471 | 0.0533 | 0.0431 |
|  | Stein | 0.0452 | **0.0151** | 0.0176 | 0.0196 | 0.0214 | 0.0214 | 0.0183 |
| III |  |  |  |  |  |  |  |  |
| $n = 10$ | RMSE($d_E$) | 0.0338 | 0.0333 | 0.0336 | 0.0335 | 0.0333 | **0.0331** | 0.0336 |
|  | RMSE($d_S$) | 0.0195 | 0.0193 | 0.0194 | 0.0194 | 0.0192 | **0.0191** | 0.0194 |
|  | Stein | 0.0017 | **0.0016** | **0.0016** | **0.0016** | **0.0016** | **0.0016** | **0.0016** |
| $n = 30$ | RMSE($d_E$) | 0.0329 | 0.0324 | 0.0327 | 0.0327 | 0.0324 | **0.0322** | 0.0328 |
|  | RMSE($d_S$) | 0.0187 | 0.0184 | 0.0185 | 0.0185 | 0.0183 | **0.0182** | 0.0185 |
|  | Stein | 0.0015 | 0.0015 | 0.0015 | 0.0015 | **0.0014** | **0.0014** | 0.0015 |
| IV |  |  |  |  |  |  |  |  |
| $n = 10$ | RMSE($d_E$) | 0.119 | 0.1012 | 0.1006 | **0.0991** | 0.0996 | 0.109 | 0.1049 |
|  | RMSE($d_S$) | 0.1202 | 0.082 | 0.0818 | **0.0811** | 0.0822 | 0.086 | 0.0922 |
|  | Stein | 0.1503 | 0.064 | 0.0637 | 0.0639 | 0.0676 | **0.0636** | 0.0639 |
| $n = 30$ | RMSE($d_E$) | 0.081 | 0.0618 | 0.0598 | **0.0582** | 0.0618 | 0.0795 | 0.0643 |
|  | RMSE($d_S$) | 0.0828 | 0.0489 | **0.0469** | 0.0472 | 0.0503 | 0.0572 | 0.0528 |
|  | Stein | 0.0825 | 0.0223 | **0.021** | 0.0228 | 0.0251 | 0.0235 | 0.0217 |

TABLE 3
*Measures of efficiency, with $k = 3$ and $\sigma = 0.1$. RMSE is the root mean square error using either the Euclidean norm or the Procrustes size-and-shape norm, and "Stein" refers to the risk using the Stein loss function. The smallest value in each row is highlighted in bold. The mean has parameters $\lambda_1 = 1, \lambda_2 = 0.001, \lambda_3 = 0.001$. The error distributions for Models I–IV are Gaussian (square root), Guassian (Cholesky), log-Gaussian and Student's $t_3$, respectively*

|   |   | $\hat{\Sigma}_E$ | $\hat{\Sigma}_C$ | $\hat{\Sigma}_S$ | $\hat{\Sigma}_H$ | $\hat{\Sigma}_L$ | $\hat{\Sigma}_R$ | $\hat{\Sigma}_F$ |
|---|---|---|---|---|---|---|---|---|
| I | | | | | | | | |
| $n=10$ | RMSE($d_E$) | 0.0999 | 0.2696 | 0.0894 | **0.0876** | 0.1014 | 0.5112 | 0.092 |
| | RMSE($d_S$) | 0.2091 | 0.2172 | 0.1424 | 0.1491 | **0.1072** | 0.3345 | 0.1439 |
| | Stein | 53.4893 | 28.1505 | 25.079 | 27.7066 | **12.4056** | 15.2749 | 25.497 |
| $n=30$ | RMSE($d_E$) | 0.0708 | 0.2836 | 0.0552 | **0.0531** | 0.0801 | 0.5515 | 0.0587 |
| | RMSE($d_S$) | 0.2064 | 0.2112 | 0.1301 | 0.1388 | **0.087**v | 0.3484 | 0.1317 |
| | Stein | 53.3301 | 25.8512 | 22.2974 | 25.378 | **8.5161** | 12.95 | 22.6973 |
| II | | | | | | | | |
| $n=10$ | RMSE($d_E$) | 0.0907 | 0.4879 | 0.0844 | **0.0839** | 0.1104 | 0.75 | 0.0861 |
| | RMSE($d_S$) | 0.1669 | 0.3571 | 0.1139 | 0.1176 | **0.1023** | 0.5168 | 0.1151 |
| | Stein | 34.2082 | 9.8147 | 15.4552 | 16.4905 | 10.2085 | **8.6754** | 15.7207 |
| $n=30$ | RMSE($d_E$) | 0.0606 | 0.5151 | 0.0509 | **0.0504** | 0.0954 | 0.7787 | 0.0533 |
| | RMSE($d_S$) | 0.1632 | 0.3369 | 0.1022 | 0.1067 | **0.0887** | 0.5369 | 0.1035 |
| | Stein | 33.9321 | 7.6303 | 13.4332 | 14.63 | 7.9578 | **7.4431** | 13.693 |
| III | | | | | | | | |
| $n=10$ | RMSE($d_E$) | 0.0315 | 0.0312 | 0.0313 | 0.0313 | **0.0311** | 0.0251 | 0.0315 |
| | RMSE($d_S$) | 0.0162 | 0.016 | 0.0161 | 0.0161 | 0.016 | **0.013** | 0.0162 |
| | Stein | 0.0034 | 0.0029 | 0.0029 | 0.0029 | **0.0028** | **0.0028** | 0.0029 |
| $n=30$ | RMSE($d_E$) | 0.031 | 0.0307 | 0.0309 | 0.0309 | 0.0306 | **0.0244** | 0.031 |
| | RMSE($d_S$) | 0.0156 | 0.0154 | 0.0155 | 0.0155 | 0.0154 | **0.0123** | 0.0156 |
| | Stein | 0.0024 | **0.0019** | **0.0019** | **0.0019** | **0.0019** | **0.0019** | **0.0019** |
| IV | | | | | | | | |
| $n=10$ | RMSE($d_E$) | 0.1055 | 0.2519 | 0.0848 | **0.0819** | 0.0895 | 0.5214 | 0.0933 |
| | RMSE($d_S$) | 0.2187 | 0.197 | 0.1253 | 0.1301 | **0.083** | 0.3348 | 0.1317 |
| | Stein | 56.1488 | 19.7674 | 18.9143 | 20.7028 | **6.5634** | 7.875 | 17.4669 |
| $n=30$ | RMSE($d_E$) | 0.0755 | 0.2628 | 0.0523 | **0.0489** | 0.0682 | 0.5552 | 0.0616 |
| | RMSE($d_S$) | 0.2098 | 0.186 | 0.1089 | 0.1161 | **0.0635** | 0.3455 | 0.1106 |
| | Stein | 53.9159 | 16.9026 | 15.701 | 17.9492 | **4.0551** | 6.541 | 14.9515 |

In Table 3 we now consider the case $\lambda_1 = 1, \lambda_2 = 0.001, \lambda_3 = 0.001$, where $\Omega$ is close to being deficient in rank. It is noticeable that the estimators $\hat{\Sigma}_C$ and $\Sigma_R$ can behave quite poorly in this example, when using $RMSE(d_E)$ or $RMSE(d_S)$ for assessment. This is particularly noticeable in the simulations for models I, II and IV. The better estimators are generally $\hat{\Sigma}_H, \hat{\Sigma}_S$ and $\hat{\Sigma}_L$, with $\hat{\Sigma}_E$ a little inferior.

Overall, in these and other simulations $\hat{\Sigma}_H, \hat{\Sigma}_S$ and $\hat{\Sigma}_L$ have performed consistently well.

**7. Discussion.** In this paper we have introduced new methods and reviewed recent developments for estimating a mean covariance matrix where the data are covariance matrices. Such a situation appears to be increasingly common in applications.

Another possible metric is the power Euclidean metric

$$d_A(S_1, S_2) = \frac{1}{\alpha}\|S_1^\alpha - S_2^\alpha\|, \tag{10}$$

where $S^\alpha = U\Lambda^\alpha U^{\mathrm{T}}$. We have considered $\alpha \in \{1/2, 1\}$ earlier. As $\alpha \to 0$, the metric approaches the log-Euclidean metric. We could consider any nonzero $\alpha \in \mathbf{R}$ depending on the situation, and the estimate of the covariance matrix would be

$$\hat{\Sigma}_A = (\hat{\Delta}_A)^{1/\alpha}, \qquad \text{where } \hat{\Delta}_A = \arg\inf_{\Delta}\left\{\sum_{i=1}^n \|S_i^\alpha - \Delta\|^2\right\} = \frac{1}{n}\sum_{i=1}^n S_i^\alpha.$$

For positive $\alpha$ the estimators become more resistant to outliers for smaller $\alpha$, and for larger $\alpha$ the estimators become less resistant to outliers. For negative $\alpha$ one is working with powers of the inverse covariance matrix. Also, one could include the Procrustes registration if required. The resulting fractional anisotropy measure using the power metric (10) is given by

$$FA(\alpha) = \left\{\frac{k}{k-1}\sum_{i=1}^k (\lambda_i^\alpha - \overline{\lambda^\alpha})^2 \Big/ \sum_{i=1}^k \lambda_i^{2\alpha}\right\}^{1/2},$$

and $\overline{\lambda^\alpha} = \frac{1}{k}\sum_{i=1}^k \lambda_i^\alpha$. A practical visualization tool is to vary $\alpha$ in order for a neurologist to help interpret the white fiber tracts in the images.

We have provided some new methods for estimation of covariance matrices which are themselves rooted in statistical shape analysis. Making this connection also means that methodology developed from covariance matrix analysis could also be useful for applications in shape analysis. There is much current interest in high-dimensional covariance matrices [cf. Bickel and Levine (2008)], where $k \gg n$. Sparsity and banding structure often are exploited to improve estimation of the covariance matrix or its inverse. Making connections with the large amount of activity in this field should also lead to new insights in high-dimensional shape analysis [e.g., see Dryden (2005)].

Note that the methods of this paper also have potential applications in many areas, including modeling longitudinal data. For example, Cholesky decompositions are frequently used for modeling longitudinal data, both with Bayesian and random effect models [e.g., see Daniels and Kass (2001);

Chen and Dunson (2003); Pourahmadi (2007)]. The Procrustes size-and-shape metric and matrix square root metric provide a further opportunity for modeling, and may have advantages in some applications, for example, in cases where the covariance matrices are close to being deficient in rank. Further applications where deficient rank matrices occur are structure tensors in computer vision. The Procrustes approach is particularly well suited to such deficient rank applications, for example, with structure tensors associated with surfaces in an image. Other application areas include the averaging of affine transformations [Alexa (2002); Aljabar et al. (2008)] in computer graphics and medical imaging. Also the methodology could be useful in computational Bayesian inference for covariance matrices using Markov chain Monte Carlo output. One wishes to estimate the posterior mean and other summary statistics from the output, and that the methods of this paper will often be more appropriate than the usual Euclidean distance calculations.

**Acknowledgments.** We wish to thank the anonymous reviewers and Huiling Le for their helpful comments. We are grateful to Paul Morgan (Medical University of South Carolina) for providing the brain data, and to Bai Li, Dorothee Auer, Christopher Tench and Stamatis Sotiropoulos, from the EU funded CMIAG Centre at the University of Nottingham, for discussions related to this work.

I. L. Dryden
Department of Statistics
LeConte College
University of South Carolina
Columbia, South Carolina 29208
USA
and
School of Mathematical Sciences
University of Nottingham
University Park
Nottingham, NG7 2RD
UK
E-mail: ian.dryden@nottingham.ac.uk

A. Koloydenko
Department of Mathematics
Royal Holloway, University of London
Egham, TW20 0EX
UK
E-mail: alexey.koloydenko@rhul.ac.uk

D. Zhou
School of Mathematical Sciences
University of Nottingham
University Park
Nottingham, NG7 2RD
UK
E-mail: pmxdz@nottingham.ac.uk